\documentstyle[12pt]{article}
\begin{document}
\baselineskip = 24pt

\begin{titlepage}
\nonumber
\vspace{1cm}
\begin{center}{\large {\bf Phase space geometry for 
constrained lagrangian systems}}\\
\vspace{1.5cm}
V.P. Pavlov \\
\vspace{0.5 cm}
{\it Steklov Mathematical Institute,} \\
{\it Vavilov 42, GSP-1, 117966, Moscow, Russia}\\
\vspace{1.5 cm}
 A.O.Starinets \footnote[1]{E-mail: aos2839@scires.nyu.edu}\\
\vspace{0.5 cm}
{\it Department of Physics, New York University,} \\
{\it 4 Washington Pl., New York, NY, 10003-6621} \\
\vspace{2.5cm}

\begin{abstract}
We study geometry of the phase space for finite dimensional dynamical systems with degenerate Lagrangians. The Lagrangian and Hamiltonian constraint formalisms are treated as the different local-coordinate pictures of the same invariant procedure. The invariant description is given in terms of geometrical objects associated with the structure of foliation on the phase space.
\end{abstract}

\end{center}
\end{titlepage}

\section*{I. INTRODUCTION} Since the appearance of the fundamental work of
Dirac \cite{Dirac}, a certain progress has been made in understanding
geometry which underlies  constrained dynamics \cite{Faddeev} --
\cite{T}. For a system with regular (nondegenerate) Lagrangian an
appropriate phase space in the  case of Lagrangian (Hamiltonian) formalism
is the tangent bundle $TQ$ (cotangent bundle $T^*Q$) of the
configurational space $Q$. In consideration of degenerate Lagrangians it
is traditionally assumed that the phase space of the system is a subspace
of $TQ$ (Lagrangian formalism) or $T^*Q$ (Hamiltonian formalism). Tangent
and cotangent bundles are treated separatively within the framework of the
constrained dynamics, being related by the standard Legendre transformation. In this paper another point of view is advocated. 

Given a Lagrangian $l\in C^{\infty}(TQ)$, the action $S$ can be written as
a functional \begin{equation} S[s] = \int_X s^*\varphi \label{1}
\end{equation} acting on the sections $s$ of the bundle $TQ\bigotimes R$
$\rightarrow$ $R$, where $X\in R$ is a compact set, $\varphi =
-Edt+\theta_l$, $\theta_l =\frac{\partial l}{\partial v}dq$ is Cartan -
Liouville one-form and $E$ is the energy of the system. The Legendre
transformation $L$: $TQ\rightarrow T^*Q$ is a diffeomorphism for
nondegenerate $l$, so the transition to the Hamiltonian formalism can be
viewed as a change of variables in (1), $s$ being the section of $T^*Q\bigotimes R$ $\rightarrow$ $R$, \ \ $\varphi = -hdt+\theta$, where
$\theta = pdq$, $\theta_l = L^*\theta$ and $h$ is the Hamiltonian,
$E=L^*h$. 

For degenerate Lagrangians the map $L$ is in general neither surjection
nor injection. In this case the pleasant symmetry between the Lagrangian and
Hamiltonian formalisms is broken, since the image of $TQ$ by $L$ now is
only a subspace of the cotangent bundle. If one wishes to proceed in the
usual way, the machinery of Dirac - Bergmann algorithm \cite{Dirac,Bergmann} is applicable either in local-coordinate approach or
by using the invariant (coordinate-free) description in the spirit of
Gotay - Nester's work
 \cite{Gotay1, Gotay2, Gotay3}. 

The invariant description of the constrained Lagrangian dynamics begins
with the basic notion of foliation on the tangent bundle and should in
principle end up with the well-defined procedure of the reduction of the
system to the submanifold of the lower dimension (which itself must admit
a tangent bundle structure) with no second class constraints left in the
theory. The next step is to compare the result with that one of the
reduction procedure in the Hamiltonian formalism and check that the
reduced submanifolds are connected by the nondegenerate Legendre
transformation. For the specific choice of the Lagrangian function the
reduction procedure of that kind has been described in \cite{Car2} by
making use of the tangent bundle geometry formalism developed earlier in
\cite{Crampin}. 

The Hamiltonian reduction is controlled by the standard symplectic
structure on $T^*Q$ given by the closed nondegenerate two-form $\omega
=dp\wedge dq$, whereas the analog of $\omega$ in the Lagrangian formalism,
$\omega_l = L^*\omega$, is degenerate for  degenerate $\ell$. 

The usual treatment of Lagrangian dynamics starts from the invariant form
of the Euler - Lagrange equations on $TQ$, \begin{equation} \dot{f} =
X_{E}f,\label{2} \end{equation} $f\in C^{\infty}(TQ)$, where for any $g\in
C^{\infty}(TQ)$ $X_{g}$ is defined by \begin{equation}
i_{X_{g}}\omega_{\ell} = -dg.\label{3} \end{equation} One may note that
the relation (3) also plays the key role in geometric quantization
\cite{Kirillov, W}.

In this paper we pay more attention to the action (1) itself. In the case of degenerate Lagrangians the bundle $TQ\bigotimes R\rightarrow R$ appears not to be an appropriate phase space and should be replaced by the bundle $W\bigotimes R\rightarrow R$, where $W=TQ\bigoplus T^*Q$ is a Whitney sum 
of $TQ$ and $T^*Q$\footnote{After this work was completed, we learned that similar approach was previously considered (in local coordinates) by 
W. Kundt \cite{kundt} and by R. Skinner and R. Rusk (using coordinate-free formulation)\cite{rusk}} . The equations which come from the variational principle can still be written in the form of (2) but the relation (3) must be deformed taking into account the structure of the foliation on the original phase space $W$. This modification allows us to include into the scheme those functions (such as the second class constraints) which are not constant along the leaves of foliation. Then the equivalence of the Lagrangian and Hamiltonian formalisms becomes apparent and the constraints in both pictures are in one-to-one correspondence with each other. For nondegenerate $\ell$ the original phase space $W$ decouples trivially into $TQ$ and $T^*Q$ at the first step of the constraint algorithm procedure.

 It follows that the most general situation one encounters in the invariant
description of the classical dynamics is that of the presymplectic
manifold rather than the symplectic one.

The organization of the paper is as follows. In Sec. II a brief review of
the invariant formulation of the Lagrangian and Hamiltonian dynamics is
given for the systems with nondegenerate Lagrangians. The degenerate case
is treated in local coordinates in Sec.III and the invariant formulation
is provided in Sec. IV. Several illustrative examples are given in Sec.V.
The Appendix contains definitions of some geometric objects used.

\section*{II. NONDEGENERATE LAGRANGIANS}
\subsection*{A. LAGRANGIAN FORMALISM}

For a given Lagrangian $l(q,\dot{q})$ the action
\begin{equation}
S[q(t)] = \int \ell (q,\dot{q})dt \label{4}
\end{equation}
with the appropriate boundary conditions can be deliberately changed to
\begin{equation}
S[q, v] = \int dt \left( \ell (q, v) -
v^i\frac{\partial\ell}{\partial v^i}+\dot{q}^i\frac{\partial\ell}{\partial v^i}\right)dt\label{5}
\end{equation}
to be considered as a functional on the tangent bundle $TQ$ of the configurational space (smooth manifold) $Q$. We denote the independent local coordinates on $TQ$ as $(q^i , v^i)$, $i=1,...N$, $N=$dim $Q$, $l(q,v)\in C^{\infty}(TQ)$.

The Lagrangian $l$ is called nondegenerate (regular) if the matrix $\Gamma_{ij} = \partial^2 l / \partial v_i \partial v_j$ has maximal rank $R=N$. Otherwise the Lagrangian is called degenerate (singular). Note that $R$ does not depend on the particular choice of the local coordinates.

The variational principle applied to (5) gives \begin{equation}
\frac{\delta S}{\delta q^i} = \frac{\partial\ell}{\partial q^i} -
\frac{\partial^2 \ell}{\partial v^k\partial q^i}v^k+\frac{\partial^2
\ell}{\partial v^k\partial q^i}\dot{q}^k - \frac{d}{dt}\left(
\frac{\partial\ell}{\partial v^i}\right)=0, \label{6} \end{equation}
\begin{equation} \frac{\delta S}{\delta v^i} = \Gamma_{ij}(\dot{q}^j -
v^j) = 0.\label{7} \end{equation}
 For nondegenerate $l$ the last equation
implies $\dot{q}^i = v^i$. Then (7) gives the Euler - Lagrange equations.
One can define a special vector field $X_E \in \cal{X}$ $(TQ)$
corresponding to (6), \begin{equation} X_E = v^i \frac{\partial}{\partial
q^i}+\Gamma^{ij}\alpha_j \frac{\partial}{\partial v^i},\label{8}
\end{equation} where \begin{equation} \alpha_j =
\frac{\partial\ell}{\partial q^i} - \frac{\partial^2 \ell}{\partial
v^j\partial q^k}v^k \end{equation} and
$\Gamma^{ik}\Gamma_{kj}=\delta_{ij}$ such that for any $f\in
C^{\infty}(TQ)$ its time derivative is given by \begin{equation} \dot{f} =
X_E f. \end{equation} In particular, for energy $E =
\frac{\partial\ell}{\partial v^i}v^i - \ell$ we have \begin{equation}
\dot{E} = X_E E =0. \end{equation} The invariant description can be given
as follows. Let $Q$ be a finite-dimensional smooth manifold, $(TQ, \pi ,
Q)$ -- its tangent bundle, $\ell\in C^{\infty}(TQ)$. The Cartan -
Liouville one-form $\theta_L\in \Lambda^1 (TM)$ is defined by
\begin{equation} \theta_L = d\ell\circ\tau , \end{equation} where $\tau$
is a canonical type $(1,1)$ tensor field on $TQ$ \cite{Vershik, Morandi}.
Then the closed nondegenerate (for nondegenerate $l$) two-form $\omega_L
=d\theta_L$ provides $TQ$ with the symplectic structure. In local
coordinates \begin{equation}\theta_L = \frac{\partial\ell}{\partial
v^i}dq^i ,\end{equation} \begin{equation} \omega_L = \Gamma_{ij}dv^i\wedge
dq^j + \frac{1}{2}M_{ij}dq^i\wedge dq^j, \end{equation} where $$ M_{ij} =
\frac{\partial^2 \ell}{\partial q^i\partial v^j}- \frac{\partial^2
\ell}{\partial v^i\partial q^j}. $$ For $f, g\in C^{\infty}(TQ)$ one can
define the Lagrangian bracket by \begin{equation} \{ f,g\}_L =
\omega_L\left( X_f,X_g \right) , \end{equation} where the vector field
$X_f\in \cal{X}$ $(TM)$is defined by \begin{equation} i_{X_f}\omega_L =
-df.\label{g} \end{equation} Since $\omega_L$ is nondegenerate, the map
$\Omega^1(TQ)\rightarrow\cal{X}$ $(TQ)$ (16) is an isomorphism. In local
coordinates we have \begin{equation} \{ f,g\}_L = \Gamma^{ij}\left(
\frac{\partial f}{\partial v^i}\frac{\partial g}{\partial q^j} -
\frac{\partial f}{\partial q^i}\frac{\partial g}{\partial v^j}\right)-
M^{ij}\frac{\partial f}{\partial v^i}\frac{\partial g}{\partial v^j},
\end{equation} where $\Gamma^{ik}\Gamma_{kj}=\delta^{i}_{j}$, $M^{ik} =
\Gamma^{ik}M_{kl}\Gamma^{lj}$. This bracket is antysimmetric. It obeys
Jacobi identity since $d\omega_L =0$. The energy $E$ can be introduced as
a smooth function on $TQ$ by \begin{equation} E = i_Z\theta_L -\ell,
\end{equation} where $Z=\left\{ X\in \cal{X}(TQ):\tau X =\triangle
\right\}$, $\triangle$ is the Liouville vector field. 

Putting all this together, we can write the action (5) in the invariant form (1). For the energy (18) one can define the corresponding vector field using (16). A simple local-coordinate calculation shows that this vector field is equivalent to the field $X_E$ (8) which originates from the variational principle. Therefore for
any $ f\in C^{\infty}(TQ)$ we can write \begin{equation} \dot{f} = X_E f =
df(X_E) = \omega_L\left( X_E,X_f\right) = \{ E, f\}_L . \end{equation} In
particular, \begin{equation} \dot{E} = dE(X_E) = 0. \end{equation}

\subsubsection*{B. Hamiltonian formalism}
The transition to the Hamiltonian form of dynamics is provided by the Legendre transformation 
 $L: T_q Q\rightarrow T_{q}^*Q$ 
\begin{equation}
(q,v)\rightarrow (q,p),\hspace{0.5 cm} p_i = \frac{\partial\ell}{\partial
v^i},
\end{equation}
$(q,p)\in T^*Q$ which is a diffeomorphism for nondegenerate $l$. In the new variables the action (1) becomes
$$
S[q,v] \rightarrow S[q,p],
$$
\begin{equation}
S = \int_X s^*\varphi , 
\end{equation}
where $\varphi = -hdt +\theta$,
$\varphi \in\Omega^1 (T^*Q\bigotimes R)$, $X\subset R^1$,
$h(q,p)=(L^*)^{-1}(E)$ is the Hamiltonian, $\theta =p_idq^i$ is the canonical one-form on $T^*Q$, $\theta_L = L^*(\theta )$.

The variational principle applied to (22) gives the standard Hamilton's equations which can be associated with the vector field 
$X_h \in \cal{X}$ $(T^*Q)$(the counterpart of $X_E$ (8)),
\begin{equation}
X_h = \frac{\partial h}{\partial
p^i}\frac{\partial}{\partial q^i} - \frac{\partial
h}{\partial q^i}\frac{\partial}{\partial p^i}.
\end{equation}
The field $X_h$ can be defined invariantly by 
\begin{equation}
i_{X_h}\omega = -dh,\label{h}
\end{equation}
where $\omega =d\theta$ is the nondegenerate two-form which provides $T^*Q$ with the standard symplectic structure. Then  $\forall f\in C^{\infty}(T^*M)$ we have
\begin{equation}
\dot{f} = X_H f = df(X_H) = \omega\left( X_H,
X_f\right) = \{ H, f\},
\end{equation}
where  $\{ f,g\}$ is the Poisson bracket. In particular,
\begin{equation}
\dot{h} = dh(X_h) = 0.
\end{equation}

\section*{III. DEGENERATE LAGRANGIANS}

For degenerate Lagrangians the Legendre transformation (21) is not a bijection. This means in particular that $R$= rank
$\Gamma_{ij}$ $\leq$ $N$ so the condition $\dot{x}^i =v^i$ 
is valid now only for a part of coordinates, i.e. for $\dot{x}^a =v^a$, a=1,...$R$. The transition to the Hamiltonian formulation can no longer be considered as a change of variables in the action (1). To cure this situation one should extend the phase space by adding $N-R$ new independent variables  $p_{\mu}$($\mu$=1,...$N-R$) and write the action $S$ as follows \cite{Pavlov}
\begin{equation}
S[q,v,p_{\mu}] = \int dt \left( \ell (q,v) - v^a
\frac{\partial\ell}{\partial v^a} + \dot{q}^a
\frac{\partial\ell}{\partial v^a} +p_{\mu}(\dot{q}^{\mu}-
v^{\mu})\right)\label{ol}
\end{equation}
which can also be presented in the form
\begin{equation}
S = \int \varphi_1 ,\hspace{1cm} 
\end{equation}
with $\varphi_1 = -E_1dt +
\theta_1$, 
where 
\begin{equation}
E_1 = v^a \frac{\partial\ell}{\partial v^a} +
p_{\mu}v^{\mu} -\ell , 
\end{equation}
\begin{equation}
 \theta_1 =
\frac{\partial\ell}{\partial v^a}dq^a + p_{\mu}dq^{\mu
}.\end{equation}
The variation of $S$ gives the equations
\begin{equation}
\dot{v}^a = \Gamma^{ab} \frac{\partial\ell}{\partial q^b}-
\Gamma^{ab} \frac{\partial^2 \ell}{\partial v^b\partial
q^i}v^i - \Gamma^{ab}\Gamma_{b\mu}\dot{v}^{\mu},
\end{equation}
\begin{equation}
\dot{p}_{\mu} = \frac{\partial\ell}{\partial q^{\mu}},
\end{equation}
\begin{equation}
\dot{q}^i - v^i=0,
\end{equation}
\begin{equation}
p_{\mu} - \frac{\partial\ell}{\partial v^{\mu}} = 0.
\end{equation}
The last equation is a constraint in the space of variables $q^i , v^i , p_{\mu}$. The time derivative of a smooth function $f(q^i , v^i , p_{\mu})$ can be written now as 
\begin{equation}
	\dot{f} = Y^{(1)}f + \dot{v}^{\mu}Y_{\mu}^{(1)}f, \end{equation}
\begin{equation}
	\phi_{\mu} = p_{\mu }-\frac{\partial l}{\partial v^{\mu}}=0,
\end{equation}
where
\begin{equation}
Y^{(1)} =v^i \frac{\partial}{\partial q^i} +\Gamma^{ab}
\alpha_b\frac{\partial }{\partial v^a} +
\frac{\partial\ell}{\partial q^{\mu}}\frac{\partial
}{\partial p^{\mu}},
\end{equation}
\begin{equation}
\alpha_b = \frac{\partial\ell}{\partial q^b} -
\frac{\partial^2 \ell}{\partial v^b\partial q^i}v^i ,
\end{equation}
\begin{equation}
Y_{\mu}^{(1)} = b^i_{\mu} \frac{\partial}{\partial v^i},
\end{equation}
and $b^{i}_{\mu}$ are the null-vectors of $\Gamma_{ij}$.

The action (28), however, does not have an invariant form. For the sake of invariance one should write $S$ as
\begin{equation}
S[q,v,p] = \int dt\left( l-p_iv^i + p_i\dot{q}^i \right).
\end{equation}
Application of the variational principle gives dynamical equations 
\begin{equation}
\dot{p}_i = \frac{\partial l}{\partial v^i},
\end{equation}
\begin{equation}
\dot{q}^i = v^i
\end{equation}
as well as the {\it primary} constraints
\begin{equation}
\phi_i = p_i -\frac{\partial l}{\partial v^i} =0.
\end{equation}
The action (40) can be considered as the basic one for both the Lagrangian and the Hamiltonian formulations for either degenerate or nondegenerate Lagrangians.

Equations (41)-(42) imply that the evolution of any smooth function $f(q,v,p)$ is given by
\begin{equation}
	\dot{f} = Yf + \dot{v}^{i}Y_{i}f, 
\end{equation}
\begin{equation}
	\phi_{i} = p_{i}-\frac{\partial l}{\partial v^{i}}=0,
\end{equation}
where
\begin{equation}
Y =v^i \frac{\partial}{\partial q^i} +
\frac{\partial\ell}{\partial q^{i}}\frac{\partial
}{\partial p^{i}},
\end{equation}
\begin{equation}
Y_{i} = \frac{\partial}{\partial v^i}.
\end{equation}
Since $Y_i\phi_j$=-$\Gamma_{ij}$, the Dirac's compatibility condition $\dot{\phi}_i=0$ leads to determination of $R$ accelerations 
\begin{equation}
\dot{v}^a = (Y\phi_b )\Gamma^{ba} - \dot{v}^{\mu}\Gamma_{\mu b}\Gamma^{ba}
\end{equation}
as well as to the new constraints
\begin{equation}
\phi^{(1)}_{\mu }= b^{i}_{\mu }\left( Y\phi_i \right) =0.
\end{equation}
At the same time a part of constraints in (45), namely those for which
\begin{equation}
\mbox{det}(Y_a\phi_b )\neq 0,
\end{equation}
$Y_a = \frac{\partial}{\partial v^a}$, can be resolved. Because of the condition (50), we are free to choose either $v^a$ or $p_a$ as the local coordinates on the resulting surface $M_1 = \left\{ q^i, p_i , v^i : \phi_a = 0, a=1,...R\right\}$. As a result of this choice one gets either Lagrangian or Hamiltonian formulation, correspondingly.
\subsection*{A. LAGRANGIAN FORMALISM}
In the Lagrangian scheme the system (44)-(45) on $M_1$ becomes
	\begin{equation}
	\dot{f} = Y^{(1)}f + \dot{v}^{\mu}Y_{\mu}^{(1)}f, 
\end{equation}
\begin{equation}
	\phi_{\mu}\mid_{M_1} = 0,
\end{equation}
\begin{equation}
        \phi_{\mu}^{(1)}\mid_{M_1} = 0,     
\end{equation}
where $Y^{(1)}$ and $Y^{(1)}_{\mu}$ have been defined in (37)-(39). The action $S$ (40) confined to $M_1$ is given by (27). We will call the primary constraints for which (50) (and consequently (48)) holds {\it the primary constraints of the second class}. The rest of them are called the primary constraints of the first class.  

Next the condition $\dot{\phi}_{\mu}^{(1)}=0$ is to be verified. This gives $R_1$ determined accelerations,
\begin{equation}
\dot{v}^{a_1} = - \left( Y^{(1)}\phi_{b_1}^{(1)}\right)\stackrel{(1)}{\gamma}^{b_1a_1} - \dot{v}^{\mu_1}\left( Y^{(1)}_{\mu_1}\phi_{b_1}^{(1)}\right)\stackrel{(1)}{\gamma}^{b_1 a_1},
\end{equation}
and a new set of constraints,
\begin{equation}
\phi^{(2)}_{\mu_1}= b^{(2)\mu}_{\mu_1}\left( 
Y^{(1)}\phi^{(1)}_{\mu}\right)=0. \end{equation}
Here $\stackrel{(1)}{\gamma}_{\mu\nu}$ = $Y^{(1)}_{\mu}\phi_{\nu}^{(1)}$,
 $R_1$=rank $\stackrel{(1)}{\gamma}_{\mu\nu}$, $a_1$=1,...$R_1$,
 $\mu_1$=1,...$N-R-R_1$ and $b^{(2)\mu}_{\mu_1}=\delta^{\mu}_{\mu_1} -
 \delta^{\mu}_{a_1}\gamma^{a_1b_1}\gamma_{b_1\mu_1}$ are the null vectors of $ \stackrel{(1)}{\gamma}_{\mu\nu}$. Again,
 among the secondary constraints (53) there are $R_1$ constraints of the second class (det$\gamma_{ab}\neq 0$) as well as $N-R-R_1$ those of the first class. The system therefore can be reduced to the surface $M_2 \subset M_1$ of the second class secondary constraints $\phi^{(1)}_{a_1}=0$ (the corresponding number of the primary constraints ($R_1$) must also be resolved, since $\dot{\phi}_{a_1}\sim \phi_{a_1}^{(1)}$). The evolution then is given by 
\begin{equation}
	\dot{f} = Y^{(2)}f + \dot{v}^{\mu_1}Y_{\mu_1}^{(2)}f, 
\end{equation}
\begin{equation}
	\phi_{\mu_1}\mid_{M_2} = 0,
\end{equation}
\begin{equation}
        \phi_{\mu_1}^{(1)}\mid_{M_2} = 0,
\end{equation}
\begin{equation}
     \phi_{\mu_1}^{(2)}\mid_{M_2} = 0,
\end{equation}
where 
\begin{equation}
Y^{(2)}= \left( Y^{(1)} -\left( Y^{(1)}\phi^{(1)}_{b_1}\right)\stackrel{(1)}{\gamma}^{b_1 a_1}Y^{(1)}_{a_1}\right)\mid_{M_2},
\end{equation}
\begin{equation}
Y^{(2)}_{\mu_1}= b^{(2)\mu}_{\mu_1}Y^{(1)}_{\mu}\mid_{M_2}.
\end{equation}
The following iterations are straightforward. After the k-th step the system becomes
\begin{equation}
	\dot{f} = Y^{(k)}f + \dot{v}^{\mu_{k-1}}Y_{\mu_{k-1}}^{(k)}f,
\end{equation}
plus the set of constraints
\begin{equation}
     	\phi_{\mu_{k-1}}^{\alpha}\mid_{M_k} = 0,
\end{equation}
where $\alpha = 0,1,...k$, $\phi_{\mu_{k-1}}^{(0)} =\phi_{\mu_{k-1}}$,
\begin{equation}
Y^{(k)}= \left( Y^{(k-1)} -\left( Y^{(k-1)}\phi^{(k-1)}_{b_{k-1}}\right)\stackrel{(k-1)}{\gamma}^{b_{k-1} a_{k-1}}Y^{(k-1)}_{a_{k-1}}\right)\mid_{M_k},
\end{equation}
$M_k =\left\{ x\in M_{k-1}: \phi^{(\alpha)}_{a_{k-1}}(x)=0, \alpha
 = 0,...k \right\}$, $M_k\subset M_{k-1}\subset M_{k-1} \subset \cdots 
\subset W$.

The iteration process terminates, if:

1) for certain $k$ we have 
$$
R_k = N - R_1 - \cdots - R_{k-1},
$$
where $R_k$= rank $\stackrel{(k)}{\gamma}_{b_{k-1} a_{k-1}}$. In this case all the accelerations are determined and the dynamics confined to the final constraint surface $M_k$ is totally fixed.

2) all the constraints of generation $k+1$ are reducible to those of the previous generations or are identically zero. In this case $N-R_1-\cdots - R_k$ accelerations remain undetermined and the system posesses certain ``gauge freedom''.

\subsection*{B. HAMILTONIAN FORMALISM}
Let us rewrite the system (44) - (45) in a slightly different form
\begin{equation}
	\dot{f} = \left\{ E, f\right\} +\dot{v}^i Y_i f,
\end{equation} 
\begin{equation}
	\phi_{i} = 0,
\end{equation}
where $E(p,q,v) = p_iv^i -l$ and $\left\{ f,g \right\}$ is the usual Poisson bracket generated by the two-form $dp\wedge dq$ (which is of course degenerate in the space of variables $p, q, v$). As it was in (54) - (55), we determine $R$ accelerations,
\begin{equation}
\dot{v}^a = \left\{ E, \phi_b \right\}\Gamma^{ba} - \dot{v}^{\mu}\Gamma_{\mu b}\Gamma^{ba},
\end{equation}
and a set of new constraints,
\begin{equation}
\phi^{(1)}_{\mu}=b^{i}_{\mu}\left\{ E, \phi_i \right\} = 0.
\end{equation}
The condition (50) allows us to resolve a certain part of
 constraints in (66). To produce the Hamiltonian scheme, one should
 choose $p_a$ rather than $v^a$ as a set of local coordinates on the surface
 $M_1$ of the resolved constraints. Due to the identity
 $\Gamma_{\mu\nu}=\Gamma_{\mu a}\Gamma^{ab}\Gamma_{b\nu}$ the Routh's function
\begin{equation}
R(q^i, p_a, v^{\mu}) = \left( l - p_av^a \right)\mid_{M_1}
\end{equation}
is linear in $v^{\mu}$:
\begin{equation}
R = -h(q^i, p_a) +v^{\mu}\psi_{\mu}(q^i, p_a)
\end{equation}
(here $\psi_{\mu}(q^i, p_a) =\partial l /\partial v^{\mu}\mid_{M_1}$), so 
the action (40) being confined to $M_1$ in coordinates $p_a, q^i, v^{\mu}$ reads
\begin{equation}
S[q^i, p_a, v^{\mu}]=\int dt \left( -h_T + p_i\dot{q}^i\right),
\end{equation}
where 
\begin{equation}
h_T=h(q^i, p_a) +v^{\mu}\phi_{\mu},
\end{equation}
$$
\phi_{\mu}=p_{\mu} - \psi_{\mu}(q^i, p_a).
$$
Correspondingly, the system (65) - (66) becomes
\begin{equation}
	\dot{f} = \left\{ h, f\right\} + v^{\mu}\left\{\phi_{\mu}, f\right\} +\dot{v}^{\mu} Y_{\mu}f, 
\end{equation}
\begin{equation}
	\phi_{\mu}\mid_{M_1} = 0,
\end{equation}
\begin{equation}
\phi_{\mu}^{(1)}\mid_{M_1} = 0,
\end{equation}
Note that (28) and (71) are equivalent. The transition between the Lagrangian and the Hamiltonian formulations is a smooth change of variables in the action, as it was for nondegenerate Lagrangians. There is no need to add voluntarily the primary constraints $\phi_{\mu}$ to the function $h(q^i, p_a)$, since the old Dirac's construction $h_T$ naturally appears in (71) and in (73).

The condition $\dot{\phi}^{(1)}_{\mu}$=$0$ leads to the determination of some accelerations,
\begin{equation}
\dot{v}^{a_1} =- \left\{ h_T, \phi^{(1)}_{b_1}\right\}\gamma^{b_1 a_1} -
 \dot{v}^{\mu_1}\gamma_{\mu_1 b_1}\gamma^{b_1  a_1},
\end{equation}
and to the new constraints,
\begin{equation}
\phi^{(2)}_{\mu_1} = b^{(2)\mu}_{\mu_1} \left\{ h_T, \phi^{(1)}_{\mu_1}\right\}=0. 
\end{equation}
Here $\gamma_{\mu\nu}$=$\partial (\phi^{(1)}_{\nu}\mid_{M_1})/\partial v^{\mu}$. Now we can reduce the system to the surface $M_2\subset  M_1$ of the second class secondary constraints $\phi^{(1)}_{a_1}=0$. In the Hamiltonian scheme we can do this explicitly by expressing $v^{a_1}$:
\begin{equation}
v^{a_1} =- \left\{ h, \phi_{b_1}\right\}\gamma^{b_1 a_1} -
 v^{\mu_1}\left\{ \phi_{\mu_1},\phi_{b_1}\right\}\gamma^{b_1  a_1}.
\end{equation}
Then the system (73)-(75) on $M_2$ can be written as
\begin{equation}
	\dot{f} = \left\{ h, f\right\}^{*}_1  + v^{\mu}\left\{\phi_{\mu_1}, f\right\}^{*}_1 +\dot{v}^{\mu_1} Y_{\mu_1}f, 
\end{equation}
\begin{equation}
	\phi_{\mu_1}\mid_{M_2} = 0,
\end{equation}
\begin{equation}
\phi_{\mu_1}^{(1)}\mid_{M_2} = 0,
\end{equation}
\begin{equation}
 \phi_{\mu_1}^{(2)}\mid_{M_2} = 0
	\end{equation}
where 
\begin{equation}
\left\{ f,g\right\}^*_1=\{ f,g\}-\{ f,\phi_{b_1}\}\gamma^{b_1
a_1}\{ \phi_{a_1} ,g\}
\end{equation}
is the Dirac bracket in respect to the second class primary constraints. Next iteration gives us another set of the determined accelerations,
     
\begin{equation}
\dot{v}^{a_2} =- \left\{ h, \phi^{(2)}_{b_2}\right\}^*_1\stackrel{(2)}{\gamma}^{b_2 a_2} - v^{\nu_1}\left\{ \phi_{\nu_1}, \phi^{(2)}_{b_2}\right\}^*_1\stackrel{(2)}{\gamma}^{b_2 a_2}-
 \dot{v}^{\mu_2}\stackrel{(2)}{\gamma}_{\mu_2 b_2}\gamma^{b_2  a_2},
\end{equation}
velocities,
\begin{equation}
v^{a_2} =- \left\{ h, \phi^{(1)}_{b_2}\right\}^*_1\stackrel{(2)}{\gamma}^{b_2 a_2} -
 v^{\mu_2}\left\{ \phi_{\mu_2},\phi_{b_2}^{(1)}\right\}^*_1\stackrel{(2)}{\gamma}^{b_2  a_2},
\end{equation}
where 
\begin{equation}
\stackrel{(2)}{\gamma}_{\mu_1\nu_1}=\partial \phi_{\nu_1}^{(2)}/\partial v^{\mu_1} = \left\{ \phi_{\mu_1},\phi_{\nu_1}^{(1)}\right\}^*_1,
\end{equation}
and constraints
\begin{equation}
\phi^{(3)}_{\mu_2} = b^{(3)\mu_1}_{\mu_2}\left( \left\{ h, \phi^{(2)}_{\mu_1}\right\}^*_1 + v^{\nu_1} \left\{ \phi_{\nu_1}, \phi^{(2)}_{\mu_1}\right\}^*_1\right) =0. 
\end{equation}
The system (79)-(82) then becomes 
\begin{equation}
	\dot{f} = \left\{ h, f\right\}^{*}_2  + v^{\mu_2}\left\{\phi_{\mu_2}, f\right\}^{*}_2 +\dot{v}^{\mu_2} Y_{\mu_2}f, 
\end{equation}
\begin{equation}
	\phi_{\mu_2}\mid_{M_3} = 0,
\end{equation}
\begin{equation}
\phi_{\mu_2}^{\alpha}\mid_{M_3} = 0, \alpha =1,2,3,
\end{equation}
where
\begin{equation}
\left\{ f,g\right\}^*_2=\left\{ f,g\right\}^*_1-\left\{ f,\phi_{b_2}^{(1)}\right\}^*_1\stackrel{(2)}{\gamma}^{b_2
a_2}\left\{ \phi_{a_2} ,g\right\}^*_1 + 
\left\{ g,\phi_{b_2}^{(1)}\right\}^*_1\stackrel{(2)}{\gamma}^{b_2
a_2}\left\{ \phi_{a_2} ,f\right\}^*_1 .
\end{equation}
The iterative procedure continues until either all acceleration ( and velocities) are determined (and the final phase space is a submanifold of $T^*Q$) or the constraints of the k-th generation are reducible to the previous ones (or are identically zero).

\subsection*{IV. INVARIANT FORMULATION}

Let $M$ be a real smooth $N$-dimensional manifold, $(TM, \pi_1 , M)$ 
 its tangent bundle, $(T^*M, \pi_2 , M)$ its cotangent bundle. Let $W$=$
 TM\bigoplus T^*M$ denotes a Whitney sum of these two bundles (Fig.1).

 
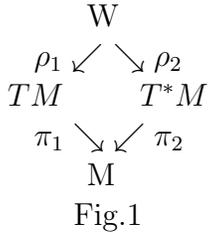
\begin{picture}(100,100)
\put (50,80){W} \put (50,20){M}
\put (20,50){$TM$} \put (70,50){$T^*M$}
\put (30,65){$\rho_1 \swarrow$}
 \put (60,65){$\searrow\rho_2$}
\put (30,35){$\pi_1\searrow$} \put (60,35){$\swarrow\pi_2$}
\put (45,5){Fig.1}
 \end{picture}

  Consider  $\rho^*_2\theta\in
\Omega^1(W)$, where $\theta$ is the canonical one-form on $T^*M$. To simplify
 notations we denote $\rho^*_2\theta$ by $\theta$ again. The two-form
 $\omega =d\theta $ is closed and degenerate on $W$, Ker $\omega =
 {\cal X}^V$(TM)$\subset {\cal X}$(W). The elements of Ker $\omega$ form
 an integrable distribution ${\cal D}$ on $W$ thus defining a foliation
 $F$ of $W$, the leaves of $F$ being the maximal integral manifolds of
 ${\cal D}$. The foliation $F$ has codimension $q=3N-$rank${\cal D}$ = dim 
\ Coker $\omega$=$2N$. We assume that the space of leaves ${\cal L}$ has
 a structure of manifold, so $F$ is a regular foliation \cite{Marmo} of
 codimension $q$. Now consider the de Rham complex of $F$. We have $\Omega^1 W/{\cal  L}$=$T^*W/{\cal L}$, $\Omega^i W/{\cal L}$=$\Lambda^i\Omega^1 W/{\cal L}$, 
 $\Omega  W/{\cal L}$=$\bigoplus_{i\geq 0} \Omega^i W/{\cal L}$. The
 first order differential operator $d_q$:
$\Omega W/{\cal L}$ $\rightarrow$ $\Omega W/{\cal L}$ has the following properties \cite{Manin}:

i) $d_q$ : $C^{\infty}(W)\rightarrow \Omega^1 W/{\cal L}$ is a composition
 $C^{\infty}(W)\stackrel{d}{\rightarrow}\Omega^1 W\rightarrow T^*W/{\cal L}$,

ii) $d_q$($\omega^p\wedge\omega^s$)= $d_q\omega^p\wedge \omega^s +(-1)^p
\omega^p\wedge d_q\omega^s$, $\omega^p\in \Omega^pW/{\cal L}$,

iii) $d_{q}^2$ =0.

Note also that $d_0=d$ : $\Omega W\rightarrow\Omega W$. In any Frobenius 
neighborhood $U\subset W$ one can choose the local coordinate system
 $(x^1,...x^d, y^{d+1}...y^{d+q})$ such that ${\cal D}_U$ is generated by
 the set $\{ \partial /\partial x^{\mu} \}$, $\mu$=1,...$d$=rank ${\cal D}$. Then the action of $d_q$ on any $\alpha\in \Omega^p W/{\cal L}$,
$$
\alpha = a_{\mu_1 ...\mu_p}(x,y)dx^1\wedge \cdots \wedge dx^{\mu_p},
$$
can be described as follows
$$
d_q \alpha = \frac{\partial a_{\mu_1 ...\mu_p}(x,y)}{\partial x^{\nu}}dx^{\nu}
\wedge dx^1\wedge \cdots \wedge dx^{\mu_p},
$$
$\mu ,\nu$=$1,...d$. In other words, provided $\left\{ K_i\right\}^{d}_{i=1}$ is a basis in ${\cal D}$ and $\left\{ \theta^i\right\}^{d}_{i=1}$
is a dual basis in ${\cal D}^*$, the action of $d_q$ is given by
$$
d_q\alpha =\left( K_i a_{i_1\cdots i_p}\right)\theta^i\wedge 
\theta^1\wedge\cdots\wedge\theta^{i_p}. $$
Let us denote $\omega^{q}_{f}=d_qf$, $f\in C^{\infty}(W)$. Now we are ready to proceed with the invariant formulation of dynamics.

Let $Z=\{ X\in {\cal X}(W): \tau (\rho_{1*}X)=\triangle \}$, where $\tau$ is the vertical endomorphism, $\triangle$ is the Liouville field. Then for any Lagrangian $l\in C^{\infty}(TM)$ the energy $E\in C^{\infty}(W)$ is defined as follows
\begin{equation}
E=i_Z\theta -\rho^{*}_1 l .
\end{equation}
The action is a functional on the smooth sections of $W\bigotimes R$$\rightarrow R$ ,
\begin{equation}
S[s] = \int_X s^*\varphi , 
\end{equation}
where $X\in R$ is a compact set, $\varphi = -Edt +\theta$, $\varphi
 \in \Omega^1 (W\bigotimes R)$. Since $W$ is only a presymplectic manifold,
 the relation (16) between one-forms and vector fields on $W$ now makes sense
 only for functions constant along the leaves of foliation $F$.
 In particular, (16) cannot be used for the second class constraints of
 any generation. However, in the local coordinate approach to the constrained
 dynamics we do not encounter such restriction. To cure this, we deform
 the correspondence (16) and put it to be
\begin{equation}
i_{X_f}\omega = -df +\omega^q_f,\label{kmkk}
\end{equation}
where $q$ is the codimension of the foliation generated by Ker $\omega$. Then
the system of equations (44) - (45) which follows from the variation of $S$ can be written in the form 
\begin{equation}
\dot{f}=df(X_E),
\end{equation}
\begin{equation}
\omega^{q}_E=0, 
\end{equation}
where $X_E$ is defined by (94). Using (94) we can rewrite (95) as
\begin{equation}
\dot{f}= \omega \left( X_E , X_f\right) +\omega^{q}_f \left( X_E \right). 
\end{equation}
Note that (94) -(95) give $\dot{E}=0$.
One can decompose $X_E$ as 
\begin{equation}
X_E = Y +K,
\end{equation}
where $K\in {\cal D}$ is an arbitrary vector. Locally, $K=\alpha^iK_i$,
 where $\{ K_i \}^{d}_{i=1}$ is a basis in ${\cal D}$. Since ${\cal D}\in 
{\cal X}^V (TW)$, the yet undetermined multipliers $\alpha^i$ are interpreted
 as accelerations $\dot{v}^i$ in any given chart. The equation (96) then reads
\begin{equation}
\omega^{q}_E = \left( K_i E\right)\theta^i = 0,
\end{equation} 
with $\theta^i\left( K_i\right) =\delta^{i}_j$. This defines a subspace
 $S\subset W$ by
\begin{equation}
\phi_i = dE\left( K_i\right) =0.
\end{equation}
For compatibility of (95) - (96) one must have therefore $X_E\in TS$, i.e.
\begin{equation}
d\phi_i \left( X_E \right) =0.
\end{equation}
If $S$ is transversal to the leaves of $F$ (in this case ${\cal D}\cap
 TS=\emptyset$ and rank $\mid d\phi_i (K_j )\mid$=$N$), the vector
 field $X_E$ is completely fixed by the condition (101) in the sense that
 all coefficients $\alpha^i$ are determined. Then we can confine the system to $S$ and choose either p's or v's as the set of local coordinates on $S$ thus generating either Hamiltonian ($S=T^*Q$) or Lagrangian ($S=TQ$) form of dynamics.

In the generic case, however, ${\cal D}\cap TS$=${\cal D}_1$
 $\neq \emptyset$, rank $\mid d\phi_i (K_j )\mid$= $R_1$ $\leq N$,
 so only $R_1$ out of $N$ accelerations are determined by (101). Instead 
of $S$ we can choose now an intermediate transverse subspace $M_1$ 
($S\subset M_1\subset W$) such that $d_1 =$dim (${\cal 
D}\cup TM_1$) = $N-R_1$. The two-form $\omega_1 =\omega\mid_{M_1}$ 
is degenerate, dim Ker $\omega_1$ = $d_1$. Now the scheme repeats itself. The system (95) -- (96) becomes  
\begin{equation}
\dot{f}=df\left(X_{E_1}\right),
\end{equation}
\begin{equation}
\omega^{q}_E\mid_{M_1}=0, 
\end{equation}
\begin{equation}
\omega^{q_1}_{E_1}\mid_{M_1}=0, 
\end{equation}
where $E_1 =E\mid_{M_1}$, $q_1 = 2N$ is a codimension of the foliation of
 $M_1$ generated by $\omega_1$. This system can be expressed equivalently in
 terms of either Lagrangian ( (51) -- (53) ) or Hamiltonian ( (73) -- (75) )
 local coordinates. A new subspace $S_1 \subset M_1$ defined by (103) 
arises here. The requirement 
$X_{E_1}$ $\subset TS_1$ gives $R_2\leq N-R_1$ determined accelerations.
 We choose $M_2$ such that $d_2 =$ dim (${\cal D}\cap TM_2$) = $N-R_1-R_2$.
 Now $\omega_2 =\omega_1 \mid_{M_2}$ is degenerate and dim Ker $\omega_2$ = $d_2$. The dynamics is governed by the system
\begin{equation}
\dot{f}=df\left(X_{E_2}\right) = \omega_2 \left( X_{E_2}, X_f\right) + \omega^{q_2}_f\left( X_{E_2}\right),
\end{equation}
\begin{equation}
\omega^{q}_E\mid_{M_2}=0, 
\end{equation}
\begin{equation}
\omega^{q_1}_{E_1}\mid_{M_2}=0,
\end{equation}
\begin{equation}
\omega^{q_2}_{E_2}\mid_{M_2}=0,
\end{equation}
The sequence stops when for certain $n$ $S_n$ is not foliated, i.e. when ${\cal D}_{n-1}\cap TM_{n-1}$ = $\emptyset$ and $\omega_n$ is nondegenerate. In this case the dynamics on $S_n = M_n\subset W$ is fixed and one has 
\begin{equation}
\dot{f}= \omega_n \left( X_{E_n}, X_f\right)
\end{equation}
for any $f\in C^{\infty}(S_n)$. As an alternative, one may discover,
 however, that for some $n$ the condition $X_{E_{n-1}}\in TS_{n-1}$ is
 automatically satisfied while $S_{n-1}$ is still foliated.
 This means that $N-R_1-\dots -R_{n-1}$ accelerations remain undetermined
 and the evolution is described by the equation
\begin{equation}
\dot{f}= \omega_n \left( X_{E_n}, X_f\right) + \omega^{q_n}_f\left( X_{E_n}\right),
\end{equation}
together with the final set of the first-class constraints
\begin{equation}
\omega^{q}_{E}\mid_{M_n}=0,\cdots \omega^{q_n}_{E_n}\mid_{M_n}=0.
\end{equation}
The invariant geometrical treatment of the first-class constraints case including gauge-fixing procedure was given by L.D. Faddeev \cite{Faddeev}.

\subsection*{V. EXAMPLES}

{\it Example 1}:
Let us consider the simplest nondegenerate Lagrangian,
\begin{equation}
l= \frac{v^2}{2}-U\left( q\right).
\end{equation}
We start with the phase space $W=TR^1\bigoplus T^*R^1$. The canonical two-form $\omega =dp\wedge dq$ is degenerate on $W$, Ker $\omega$= span$\{ \partial /\partial v\}$. The energy $E\in C^{\infty}(W)$ (92) is given by
\begin{equation}
E= pv-\frac{v^2}{2}+U\left( q\right).
\end{equation}
The primary constraint one-form $\omega_E$ can be written as 
\begin{equation}
\omega_E= d_q E = \frac{\partial E}{\partial v}dv = \phi dv =0,
\end{equation}
where $\phi = p-v$. The relation (94) gives the evolution vector field $X_E$ on $W$,
\begin{equation}
X_E= v\frac{\partial}{\partial q} - \frac{\partial U}{\partial q}\frac{\partial }{\partial p} + \dot{v} \frac{\partial}{\partial v}.
\end{equation}
Thus the dynamics on $W$ is given by $\{ \dot{f}=X_E f, \phi =0 \}$.
 The compatibility condition (101) allows to determine the acceleration
 $\dot{v}=\partial U/\partial q$. This means that $\phi$ is the
 primary constraint of the second class. Therefore, we reduce the system to $S_1$ = $\{ x\in W:\phi =0 \}$. One can choose either $(q,v)$ or $(q,p)$ as the local coordinates on $S_1$. In the first case one gets the Lagrangian scheme,
\begin{equation}
E_1= E\mid_{S_1} = \frac{v^2}{2}+U\left( q\right),
\end{equation}
\begin{equation}
X_{E_1}= v\frac{\partial}{\partial q} - \frac{\partial U}{\partial q}\frac{\partial }{\partial v}, 
\end{equation}
in the second case - the Hamiltonian description,
\begin{equation}
h= E\mid_{S_1} = \frac{p^2}{2}+U\left( q\right),
\end{equation}
\begin{equation}
X_{h}= p\frac{\partial}{\partial q} - \frac{\partial U}{\partial q}\frac{\partial }{\partial p}.
\end{equation}
In this case $W$ trivially decouples into $TQ$ and $T^*Q$.

{\it Example 2}: Consider the Lagrangian (Ref. \cite{Nesterenko})
\begin{equation}
l= \frac{v_{1}^2}{2}- v_2q_3 .
\end{equation}
The initial phase space is $W$=$TR^3\bigoplus T^*R^3$ with local coordinates $q_i, p_i, v_i$, $i=1,2,3$. The fundamental two-form $\omega = dp_i\wedge dq_i$ is degenerate on $W$, Ker $\omega$ = span $\{ \partial /\partial v_1,
\partial /\partial v_2, \partial / \partial v_3 \}$. The energy $E\in C^{\infty}(W)$,
\begin{equation}
E=p_1v_1 + p_2v_2 + p_3v_3 - \frac{1}{2}v_{1}^2 + v_2q_3,
\end{equation}
defines (by (94)) the evolution vector field $X_E =Y+K$, where
\begin{equation}
Y= v_i\frac{\partial}{\partial q_i} - v_2\frac{\partial }{\partial p_3}, 
\end{equation}
\begin{equation}
K=\dot{v}_i\frac{\partial}{\partial v_i}
\end{equation}
with yet undetermined multipliers $\dot{v}_i$. The one-form (96) of the primary constraints is given by 
\begin{equation}
\omega_E = d_q E =\frac{\partial E}{\partial v_i}dv_i =\phi_i dv_i =0,
\end{equation}
where $\phi_1 = p_1 -v_1$, $\phi_2 = p_2 +q_3$, $\phi_3 =p_3$. The compatibility condition (101) determines one of the accelerations, $\dot{v}_1 =0$, and produces two new constraints, 
\begin{equation}
\phi_{2}^{(1)} =v_3 =0, \hspace{1 cm} 
\phi_{3}^{(1)} =v_2 =0 .
\end{equation}
Thus the primary constraint $\phi_1$ is of the second class and can be resolved. On the intermediate transverse subspace $M_1 = \{ x\in W:\phi_1 =0\}$ the evolution is given by
\begin{equation}
\dot{f}= v_i\frac{\partial f}{\partial q_i} - v_2\frac{\partial f}{\partial p_3}+ \dot{v}_2\frac{\partial f}{\partial v_2} + \dot{v}_3\frac{\partial f}{\partial v_3} , 
\end{equation}
together with the set of constraints 
\begin{equation}
M_2 = \left\{ \phi_2 =0, \phi_3 =0,
 \phi^{(1)}_2 =0, \phi^{(1)}_3 =0 \right\}.
\end{equation}
 Since $\dot{\phi}^{(1)}_2 = 
\dot{v}_3 =0$, $\dot{\phi}^{(1)}_3 =\dot{v}_2 =0$, all the secondary
 constraints are of the second class and the final constrained submanifold
 $S_2 =M_2$. For any $f\in C^{\infty}(S_2)$ one has then $\dot{f} = v_1\frac{\partial f}{\partial q_1}$. The dynamics is totally fixed.

To consider the Hamiltonian form of dynamics, one should choose $p_1$ instead of $v_1$ as the local coordinate on $M_1$. Note that
\begin{equation}
E\mid_{M_1} = \frac{1}{2}p_{1}^2 + v_2q_3 +p_2v_2 +p_3v_3 = h+v_2\phi_2 +v_3\phi_3.
\end{equation}
The equation (126) can be written in the form
  \begin{equation}
\dot{f}= \left\{ h, f\right\} +v_2\left\{ \phi_2 , f\right\} +v_3\left\{ \phi_3 , f\right\} +  \dot{v}_2\frac{\partial f}{\partial v_2} + \dot{v}_3\frac{\partial f}{\partial v_3} , 
\end{equation}
together with the set $M_2$ (127). The compatibility condition again gives $\dot{v}_2 =0$, $\dot{v}_3 =0$. One can make the canonocal transformation,
$P_1 = p_1$, $Q_1 = q_1$, $P_2 =p_2$, $Q_2 = q_2 +p_3$, $P_3 = \phi_3 = p_3$,
 $Q_3 = q_3 +p_2 =\phi_2$. In terms of these new variables $S_2 = \{ P_2 =0,
 Q_2 =0, P_3 =0, Q_3 =0\}$. One has $h\mid_{S_2} = P_{1}^2$ and $\dot{f}
 = \{ h, f\}$ for any $f\in C^{\infty}(S_2)$.

{\it Example 3}: Let
$
l = \frac{1}{2} \left( q_1v_2 -q_2v_1 -q_{1}^2 - q_{2}^2\right)
$ (Ref. \cite{olga}). The two-form $\omega = dp_i\wedge dq_i$, $i=1,2$, is degenerate on $W= TR^2\bigoplus T^*R^2$, Ker $\omega$ = span $\{ \partial /\partial v_1, \partial / \partial v_2 \}$. The energy is 
\begin{equation}
E=p_1v_1 +p_2v_2 -l
\end{equation}
and the evolution vector field $X_E$ is given by 
\begin{equation}
X_E= v_1\frac{\partial }{\partial q_1} +v_2\frac{\partial }{\partial q_2} +
\left( \frac{v_2}{2} -q_1\right)\frac{\partial }{\partial p_1}-
\left( \frac{v_1}{2} -q_2\right)\frac{\partial }{\partial p_2}
+ \dot{v}_1\frac{\partial }{\partial v_1} + \dot{v}_2\frac{\partial }{\partial v_2} , 
\end{equation}
The set of primary constraints is defined by 
\begin{equation}
\omega_E = d_q E=\phi_i dv_i =0,
\end{equation}
where $\phi_1 = p_1 +q_2/2$, $\phi_2= p_2 -q_1/2$. The condition (101) gives
 only the new constraints, $\phi^{(1)}_1 = v_2 -q_1 =0$, $\phi^{(1)}_2 = v_1
 +q_2 =0$. The next iteration $\left( \dot{\phi}^{(1)}_{1}=0, 
 \dot{\phi}^{(1)}_{2}=0 \right)$, however, allows to determine $\dot{v}_1
 =-v_2$ and $\dot{v}_2 =v_1$. Therefore, the final constrained submanifold
 is $$S=\{ \phi_1 =0, \phi_2 =0, \phi^{(1)}_{1} =0, \phi^{(1)}_{2}=0\}.$$ 
The dynamics on $S$ is completely fixed and $\forall f\in C^{\infty}(S)$ we have  $\dot{f} = v_1\frac{\partial f}{\partial q_1} - q_1\frac{\partial f}
{\partial v_1}.
$

The energy (130) has the form
\begin{equation}
E = \frac{1}{2}\left( q_1^2 +q_2^2\right) +v_1\phi_1 +v_2\phi_2 .
\end{equation}
Instead of (131), (132) we can write
 \begin{equation}
\dot{f}= \left\{ E, f\right\} +  \dot{v}_1\frac{\partial f}{\partial v_1} + \dot{v}_2\frac{\partial f}{\partial v_2} , 
\end{equation}
\begin{equation}
\phi_1 =0,
\end{equation}
\begin{equation}
\phi_2 =0,
\end{equation}
$\forall f\in C^{\infty}(W)$.
The reduction to $S$ must be ``canonical'' in Hamiltonian formalism. The
 canonical transformation $P_1 = p_1 -q_2/2$, $Q_1 = q_1 /2 +p_2$, $P_2
 =p_2 - q_1 /2 =\phi_2$, $Q_2 = q_2 /2  +p_1 =\phi_1$ gives $S=\{ P_2 =0,
 Q_2 =0\}$ and $h=E\mid_{S} = \frac{1}{2}\left( P_1^2 +Q_1^2\right)$. The evolution 
equation (134) on $S$ becomes $\dot{f} =\{ h, f\}$, $f\in C^{\infty}(S)$.

{\it Example 4}: Now consider the Lagrangian (Ref.\cite{H})
$$
l = \frac{1}{2}e^{q_2}v_{1}^{2} .
$$
As it was in the previous example, $W=TR^2\bigoplus T^*R^2$, Ker $\omega$ =
 span $\{ \partial /\partial v_1 , \partial /\partial v_2 \}$. The energy 
\begin{equation}
E=p_1v_1 +p_2v_2 - \frac{1}{2}e^{q_2}v_{1}^{2}
\end{equation}
produces (by (94)) the evolution vector field 
\begin{equation}
X_E= v_1\frac{\partial }{\partial q_1} +v_2\frac{\partial }{\partial q_2} +
      \frac{1}{2}e^{q_2}v_{1}^{2}\frac{\partial }{\partial p_2}
+ \dot{v}_1\frac{\partial }{\partial v_1} + \dot{v}_2\frac{\partial }
{\partial v_2} 
\end{equation}
and the primary constrained form  
\begin{equation}
\omega_E = \phi_1 dv_1 + \phi_2 dv_2 =0,
\end{equation}
where $\phi_1 = p_1 - e^{q_2}v_1$, $\phi_2 =p_2$. The condition (101) gives $\dot{v}_1 =-v_1v_2$ as well as the new constraint $\phi^{(1)}_2 =v_1 =0$. Thus the primary constraint $\phi_1$ is of the second class. On $M_1 = \{ x\in W:\phi_1 =0\}$ we have  
\begin{equation}
\dot{f} =  v_1\frac{\partial f}{\partial q_1} +v_2\frac{\partial f}{\partial q_2} +\frac{1}{2}e^{q_2}v_{1}^{2}\frac{\partial }{\partial p_2}
-v_1v_2\frac{\partial f}{\partial v_1} + \dot{v}_2\frac{\partial f}{\partial v_2} , 
\end{equation}
\begin{equation}
\phi_2 =p_2 =0,
\end{equation}
\begin{equation}
\phi_2^{(1)} =v_1 =0.
\end{equation}
Since $\dot{\phi}_{2}^{(1)} \sim \phi^{(1)}_2$, the acceleration $\dot{v}_2$ remains undetermined. The constraints $\phi_2$ and $\phi_{2}^{(1)}$ are of the first class. We have also
\begin{equation}
E\mid_{M_1} = \frac{1}{2}p_1^2e^{-q_2} + p_2v_2 = h +v_2\phi_2,
\end{equation}
and the system (140) -(142) in Hamiltonian form reads
\begin{equation}
\dot{f}= \left\{ h, f\right\} + v_2\left\{ \phi_2 , f\right\}+
 \dot{v}_2\frac{\partial f}{\partial v_2} , 
\end{equation}
\begin{equation}
\phi_2 = p_2 =0,
\end{equation}
\begin{equation}
\phi_2^{(1)}=p_1 =0 .
\end{equation}
Once again, the acceleration $\dot{v}_2$ remains arbitrary.

{\it Example 5}: Finally, consider the Lagrangian (Ref. \cite{sundermeyer})
$$
l= \frac{1}{2}v_1^2 +q_2v_1 +(1-\alpha )q_1v_2 +\frac{\beta}{2}(q_1 -q_2)^2
$$
The form $\omega = dp_i\wedge dq_i$ is degenerate on  $W=TR^2\bigoplus T^*R^2$,
 Ker $\omega$ = span $\{ \partial /\partial v_1 , \partial /\partial v_2 \}$. The energy 
\begin{equation}
E=p_1v_1 +p_2v_2 - l
\end{equation}
gives  the evolution vector field 
\begin{equation}
X_E= v_1\frac{\partial }{\partial q_1} +v_2\frac{\partial }{\partial q_2} +
 \left( v_2(1 -\alpha )+\beta (q_1 -q_2)\right)\frac{\partial }{\partial p_1}
+\left( v_1 -\beta (q_1 -q_2)\right)\frac{\partial }{\partial p_2}
+ \dot{v}_1\frac{\partial }{\partial v_1} + \dot{v}_2\frac{\partial }{\partial v_2} 
\end{equation}
as well as  the primary constraint form  
\begin{equation}
\omega_E = \phi_1 dv_1 + \phi_2 dv_2,
\end{equation}
where $\phi_1 = p_1 -q_2 -v_1$, $\phi_2 = p_2 -(1-\alpha)q_1$. For simplicity we consider here only some particular cases.

Case A: $\alpha =0, \beta =0$.

The condition (101) gives $\dot{v}_1 =0$, so $\phi_1$ is the primary constraint of the second class. On $M_1 = \{ x\in W: \phi_1 =0\}$ we get 
\begin{equation}
X_{E_1}= v_1\frac{\partial }{\partial q_1} +v_2\frac{\partial }{\partial q_2} +
 v_1\frac{\partial }{\partial p_2}
 + \dot{v}_2\frac{\partial }{\partial v_2}. 
\end{equation}
The acceleration $\dot{v}_2$ cannot be determined. The dynamics is controlled by 
\begin{equation}
\dot{f} =X_{E_1}f,
\end{equation}
\begin{equation}
\phi_2 = p_2 -q_1 =0 .
\end{equation}
Since 
\begin{equation}
E\mid_{M_1} = \left( v_1 +q_2\right) v_1 +p_2v_2 -l = \frac{1}{2}\left( p_1 -q_2\right)^2 +v_2\phi_2 = h +v_2\phi_2,
\end{equation}
the Hamiltonian analog of (151) -(152) is 
\begin{equation}
\dot{f}= \left\{ h, f\right\} + v_2\left\{ \phi_2 , f\right\}+
 \dot{v}_2\frac{\partial f}{\partial v_2} , 
\end{equation}
\begin{equation}
\phi_2 = p_2 - q_1 =0 .
\end{equation}

Case B: $\alpha =0, \beta \neq 0$.

In this case the compatibility condition allows to determine $\dot{v}_1 =\beta (q_1 - q_2)$
 and gives the secondary constraint $\phi^{(1)}_{2}= q_2 -q_1$.
 The primary constraint $\phi_1$ is of the second class. On $M_1 = \{ x\in W:
 \phi_1 =0 \}$ the evolution is given by 
\begin{equation}
\dot{f}= v_1\frac{\partial f}{\partial q_1} +v_2\frac{\partial f}{\partial q_2} +
 \left( v_1 -\beta(q_1 -q_2) \right)\frac{\partial f}{\partial p_2}
+ \beta (q_1 -q_2 )\frac{\partial f}{\partial v_1} + \dot{v}_2\frac{\partial f}{\partial v_2}, 
\end{equation}
\begin{equation}
\phi_2 = p_2 - q_1 =0 ,
\end{equation}
\begin{equation}
\phi^{(1)}_2 = q_2 - q_1 =0 .
\end{equation}
The next iteration produces $\phi^{(2)}_{2} =v_2 -v_1$ and the next one gives $\dot{v}_2 =\beta (q_1 -q_2)$.
 As a result, on $S_3\subset M_1$, $S_3 = \{ x\in M_1 : \phi_2 =0, 
 \phi^{(1)}_{2} =0,  \phi^{(2)}_{2} =0\}$, we have $\dot{f} =
 v_1\frac{\partial f}{\partial q_1}$ and $E\mid_{S_3} = \frac{1}{2}v_1^2$. 
To work out the Hamiltonian scheme, note that 
\begin{equation}
E\mid_{M_1} =\frac{1}{2}\left( p_1 -q_2\right)^2 - \frac{\beta}{2}\left( 
q_1 - q_2\right)^2 +v_2\phi_2 = h+v_2\phi_2 .
\end{equation}
The equation (145) then reads
\begin{equation}
\dot{f} = \left\{ h, f\right\} +v_2\left\{ \phi_2 , f\right\} + 
\dot{v}_2\frac{\partial f}{\partial v_2}.
\end{equation}
Again, $\dot{\phi}^{(1)}_{2} = v_2 -v_1 = \phi^{(2)}_{2}$ and 
$\dot{\phi}^{(2)}_{2} =\dot{v}_2 =0$. Since $v_2 =-\{h,\phi^{(1)}_{2}\}$, 
we can write (149) as 
\begin{equation}
\dot{f} = \left\{ h, f\right\} - \left\{ h, \phi^{(1)}_{2}\right\}\left\{ 
\phi_2, f\right\} + \left\{ h, \phi_2\right\}\left\{ \phi^{(1)}_{2}, 
f\right\}
\end{equation}
(we have added the last term "by hand" to antisymmetrize the bracket, 
since $\{ h, \phi_2 \}$ $\sim$ $\phi^{(1)}_{2}=0$ on the surface of the 
second class constraints. The canonical transformation of the form $P_1 = 
p_1+p_2-q_1-q_2$, $Q_1 =q_1$, $P_2 =p_2 -q_1 =\phi_2$, $Q_2 =q_2 -q_1 
=\phi^{(1)}_{2}$ allows to represent the dynamics on $S_3 = \{ P_2 =0, 
Q_2 =0, \phi^{(2)}_{2} =0 \}$ in a simple way, $\dot{f} =\{ h, f\}_{P_1, 
Q_1}$, where $h=P_1^2/2$.

\subsection*{ACKNOWLEDGMENTS} 
A.S. gratefully acknowledges support from Russian Foundation of Fundamental Research. 

\subsection*{APPENDIX: THE GLOSSARY}

For convenience of the reader we give some notations and definitions of 
geometric objects used throughout the paper. More detailed information 
can be found in Ref. \cite{Marmo, Morandi}.

 \begin{enumerate}
  \item $M$ (or $Q$) denotes finite-dimensional real smooth manifold, 
$(TM, \pi_1 , M)$ --  its tangent bundle with local coordinates $\{ q^i, 
v^i\}$,  $(T^{*}M, \pi_2 , M)$ -- its cotangent bundle with 
local coordinates $\{ q^i, p_i\}$, $i=1,\dots N$ =dim $M$. 
\item ${\cal X}$(M) denotes the Lie algebra of smooth vector fields on 
$M$. Locally  $X\in {\cal X}(M)$  is $X=X^i(q)\frac{\partial}{\partial 
q^i}$, $X^i(q) \in C^{\infty}(M)$.
 \item Correspondingly, $Y\in  {\cal X}(TM)$ in local coordinates 
reads $Y=Y^i(q,v)\frac{\partial}{\partial
q^i}+V^i(q,v)\frac{\partial}{\partial v^i}$, where $Y^i$ and $V^i$ are 
smooth functions on $TM$.
\item ${\cal X}^V(TM)$=$\{ X\in {\cal 
X}(TM):\pi_* X = 0 \}$ is the algebra of vertical vector fields on $TM$. 
Any $X\in {\cal X}^V(TM)$ can be locally represented as 
$X=X^i(q,v)\frac{\partial }{\partial v^i}$.
 \item  Lift $\gamma$: ${\cal X}(M)\rightarrow {\cal X}^V(TM)$ 
maps      $X=X^i(q)\frac{\partial}{\partial q^i}$ $\in {\cal X}(M)$ to  
$X^V = X^i(q)\frac{\partial}{\partial 
v^i}\in {\cal X}^V(TM)$, where $X^V$ is the generator of 
the one-parameter group of the fiber diffeomorphisms  $t\rightarrow ( q^i, 
v^i+X^i(q)t)$.
 \item The canonical (1,1) tensor $\tau$ on $TM$ is defined as 
the composition  $\tau = \gamma\circ
\pi_*$. In coordinates we have $\tau =
dq^i\otimes\frac{\partial}{\partial v^i}$.
\item Cartan -- Liouville one-form is defined for  $f\in C^{\infty}(TM)$ 
by
 $\theta_f = df\circ\gamma$. If $\Gamma_{ij}=\frac{\partial^2 f}{\partial
v^i\partial v^j}$ has  maximal rank, $\omega_f = 
d\theta_f$  defines the symplectic structure on $TM$. 
\item    Liouville vector field $\triangle \in {\cal X}^V(TM)$ is the 
generator of the one-parameter group of dilations $(q,v) \rightarrow (q, 
e^tv)$. Locally,  $\triangle = v^i \frac{\partial}{\partial 
v^i}$.
 \item   Whitney sum  $E_1\bigoplus E_2$ of two bundles $E_1$ and $E_2$ 
with the same base $B$ is the bundle with the base $B$ and the fiber 
 $E_{1x}\bigoplus E_{2x}$   $\forall x \in B$.

\end{enumerate}


\begin{thebibliography}{99} \bibitem{Dirac} P.A.M. Dirac, Canad.J.Math. 2,
129 (1950). \bibitem{Bergmann} P.G. Bergmann, Helv.Phys.Acta Suppl.IV 79
(1956). \bibitem{Faddeev} L.D.Faddeev, Theor.Math.Phys. 1, No.1, 1 (1969).
\bibitem{Vershik} A.M. Vershik and L.D. Faddeev, Soviet Physics --Doklady,
17(1), 34 (1972). \bibitem{Gotay1} M.J. Gotay, J.M. Nester and G. Hinds,
J.Math.Phys., 19, 2388 (1978). \bibitem{Gotay2} M.J. Gotay, J.M. Nester,
Ann.Inst.H.Poincar\'{e} A30, 129 (1979). \bibitem{Gotay3} M.J. Gotay, J.M.
Nester, Ann.Inst.H.Poincar\'{e} A32, 1 (1980). \bibitem{Mukunda} G.Marmo,
N.Mukunda and J.Samuel, Rivista Nuovo Cimento 6, 1 (1983). \bibitem{Car1}
J.F. Cari\~{n}ena, J.Gomis, L.A.Ibort, and N.Rom\'{a}n, J.Math.Phys. 26,
1961 (1985). \bibitem{Pons} C.Battle, J.Gomis, J.M.Pons, and N.Rom\'{a}n,
J.Math. Phys. 27, 2953 (1986). \bibitem{Marmo} G.Marmo, E.J.Saletan,
A.Simoni, and B.Vitale, {\it Dynamical Systems, A Differential Geometric
Approach to Symmetry and Reduction} (Wiley, Chichester, 1985).
\bibitem{Car3} J.F.Cari\~{n}ena, C.Lopez,, and N. Roman-Roy, J.Math.Phys.
29, 1143 (1988). \bibitem{T} D.M. Gitman and I.V. Tyutin, {\it
"Quantization of fields with constraints"} (Springer, Berlin, 1990).
\bibitem{Nesterenko} V.V.Nesterenko and A.M.Chervyakov, Theor.Math.Phys.
64, 701 (1985). \bibitem{Pavlov} V.P.Pavlov, Theor.Math.Phys. 92, No.3,
1020 (1992). 
\bibitem{H}
M. Henneaux and C. Teitelboim, {\it Quantization of Gauge Systems} (Princeton University Press, Princeton, New Jersey, 1992).
\bibitem{sundermeyer} K.Sundermeyer, ``Constrained
dynamics'', in {\it Lecture Notes in Physics}, No.169 (Springer, Berlin,
1982). \bibitem{Kirillov} A.A.Kirillov, {\it ``Elements of the Theory of
Representations''} (Springer, New York, 1976).
\bibitem{W}
N. Woodhouse, {\it ``Geometric quantization''} (Clarendon Press, Oxford, 1980).
 \bibitem{Crampin}
M.Crampin, J.Phys.A:Math.Gen. 16, 3755 (1983). \bibitem{Car2} F. Cantrijn,
J.F. Cari\~{n}ena, M.Crampin, and L.A.Ibort, J.Geom. Phys. 3, 353 (1986).
\bibitem{Manin} Iu.I.Manin, {\it Gauge fields and complex
geometry}(Springer, Berlin, 1988). \bibitem{Morandi} G.Morandi,
C.Ferrario, G.Lo Vecchio, G.Marmo and C.Rubano, Phys.Repts. 188, 149
(1990).
 \bibitem{olga} O.Krupkov\'{a}, J.Math.Phys., 35, 6557 (1994).
\bibitem{kundt} W. Kundt, {\it Canonical Quantization of Gauge Invariant Field Theories}, Springer Tracts in Modern Physics, 40, 107 (1966). 
\bibitem{rusk} R. Skinner and R. Rusk, J.Math.Phys., 24, 2589 (1983).
\end{thebibliography}
\end{document}